\documentstyle[aps]{revtex}
\topmargin  - 0.5in
\author{Cris\'ogono R. da Silva\thanks{
Permanent Address: Departamento de F\'{\i}sica, Universidade
Federal de Alagoas}$^{\ddag}$,
Francisco A. Tamarit$^{\ddag,\S}$\\
and Evaldo M. F. Curado$^{\ddag,\P}$ \\
[0.3cm]}
\title{ INFLUENCE OF REFRACTORY PERIODS IN THE HOPFIELD MODEL
\thanks{E-mails:  {\tt crsilva@cat.cbpf.br},   {\tt
tamarit@fis.uncor.edu} and {\tt curado@cpd.unb.br}}  \\
[0.2cm]
}
\address{
$^{\ddag}$ Centro Brasileiro de Pesquisas F\'{\i}sicas \\
Rua Xavier Sigaud 150, 22290-180 Rio de Janeiro, Brazil\\[0.2cm]
$^{\S}$Fa.M.A.F., Universidad Nacional de C\'ordoba\\
Ciudad Universitaria, 5000 C\'ordoba, Argentina\\[0.2cm]
$^{\P}$International Centre of Condensed Matter Physics \\
C.P. 04667, 70919-970 Brasilia, Brazil\\}

\begin{document}

\maketitle
\begin{abstract}
We study both analytically and numerically the
effects of including refractory periods in the Hopfield model
for associative memory. These periods are introduced in the
dynamics of the network as thresholds that depend on the state
of the neuron at the previous time. Both the retrieval
properties and the dynamical behaviour are analized, and we
found that depending on the value of the thresholds and on the
the ratio $\alpha$ between the number of stored memories ($p$)
and the total number of neurons $(N$), the system presents not
only fixed points but also chaotic or ciclic orbits.
\end{abstract}

\thispagestyle{empty} 

\newpage
%

\section{Introduction}

\label{Introduction}
%

During the last years some biological features of real neurons
have been incorporated into the Hopfield model \cite{Hopfield} in
order to make it more realistic and trying to improve its
performance. Suitable modifications of the original model taking
into account biological ingredients such as thermal noise,
dilution, asymmetry, dynamical delays, among others, have been vastly
analized in the literature
\cite{Amit1,Amit2,Amit3,Derrida,Watkin,Gutfreund,Sompolinsky,Buhmann,Kleinfeld,Horn,Lopes}.
Although they usually deteriorate the retrieval
ability, it has been shown they enable the
implementation of new tasks, such as recognition of temporal
sequences\cite{Gutfreund,Sompolinsky,Buhmann,Kleinfeld,Horn,Lopes}
and categorization \cite{Fontanari,Stariolo,Silva}.

One crucial biological element originally absence in the
Hopfield model is the so called {\em refractory period}
\cite{LAmit,Schulten,Airaha}. Real neurons take about 1-2 milliseconds
to complete a cycle from the emission of a spike in the
pre-sinaptic neuron to the emission of a spike in the
post-sinaptic neuron. After this time, the neuron need again about
2 milliseconds to recover, and during this time, called {\em absolute
refractory period} (ARP), it is insensitive to afferent activiy
(i.e.,it cannot emit a second spike, no matter how large the
post-synaptic potential (PSP) may be). Following this short
ARP, the neuron enters in a new regime of about 5-7 milliseconds, in which
it partially recovers the capacity of emitting spikes, but now with a greater
excitation threshold which decreases with time. This is called
the {\em relative refractory period} (RRP). Following this
somewhat longer RRP, the threshold tends to return to its rest value and
the neuron can fire again with typical intra-network potentials.

The simplest way one can introduce these periods into the
dynamics of the Hopfield model is by means of a time
dependent threshold acting as an external field, which depends on
the recent history of the neuron. Since we want this threshold to mimic the
effects of fatigue observed in real neurons \cite{Horn}, it
should act only after the cell has emitted an electric
signal. So, we expect that the threshold depends on the mean
activity of the neuron in the previous time. The main
effect on the dynamics of the model is to introduce a tendency to
destabilize the fixed point attractors, allowing the appearance of
oscillatory behaviors. In the last years different threshold functions have
been studied\cite{Horn,Moreira,Opris,Kurten}, showing that they
enable the system to wander through the phase space, eventually
visiting different basins of attraction and simulating the
process by which the brain recognizes temporal sequences of
patterns . On the other hand, oscillating and chaotic
trajectories in the phase space seem to be more realistic than
fixed points attractors from a biological point of
view (see \cite{Airaha} and references therein).

In this work we analyze, using a mean field approach and through
numerical simulations, the behavior of the Hopfield model for associative
memory when the effect of these refractory periods are taken
into account in the dynamics of the system. Instead of
considering a fatigue like threshold function that would depend
on the large term history of the neuron \cite{Horn,Opris}, we introduce a
threshold that depends only on the state of the neuron in the previous time,
i.e., it is activated only when the neuron fires a spike. In
the section \ref{Model}, we introduce the model and describe how the
refractory periods are incorporated into its dynamics. In
section \ref{Fixed}, we obtain an equation for the
value of the superposition between the state of the system and
one of the memories (which is only valid for fixed points
dynamics), from which we can study the retrieval properties of
the model in this region. In section \ref{Numerical}, we obtain a
complete phase diagram and identified the regions of fixed points, cyclic
orbits and chaotic orbits. We have used a synchronous parallel
updating, which allows  an efficient use of modern
parallel-processing computers. Finally, in section
\ref{conclution}, we discuss the main results.

%

\section{The model}

\label{Model}
%

As in the Little \cite{Little} and Hopfield models we consider a
network of $N$ binary neurons, each one modeled by an Ising variable $%
S_i$ which take the values $\{-1,+1\}$, representing the passive and
active states, respectively. In order to take into account the effect of the
refractory period in the neuron $i$ we add a threshold that depends on the
time, but only through the value of the state $S_i(t)$ of the
neuron $i$. So the
post-synaptic potential at time $t$ is given by:
\begin{equation}
h_i(t)\;=\;h_i^H(t)-\frac \Delta 2(1+S_i(t))\;\;,
\end{equation}
where $h_i^H$ is the usual Hopfield post-synaptic potential:
\begin{equation}
\label{campo}h_i^H(t)=\sum_{j\neq i}^N\;J_{ij}\;S_j(t)\;\;.
\end{equation}
Here $J_{ij}$ is the Hopfield synaptic matrix connecting the pre and
post-synaptic neurons $j$ and $i$ and whose elements have the form:
\begin{equation}
J_{ij}=\frac 1N\;\sum_{\mu =1}^p\;\xi _i^\mu \;\xi _j^\mu \;.
\end{equation}
The $\xi_{i}^{\mu}$ are random independent variables which take the
values $\pm 1$ with the same probability and the N-bits words $\{\xi _1^\mu
,\xi _2^\mu ,\ldots ,\xi _N^\mu \}$ stand for the $p$ stored configurations (%
$\mu =1,2,\ldots ,p$). The dynamics of the network is governed by a Monte
Carlo heat bath dynamics:
\begin{equation}
S_i(t+1)=\left\{
\begin{array}{ll}
+1 & \mbox{    with probability        }\left( 1+exp(-\frac{h_i(t)}{T})\right)
^{-1} \\  &  \\
-1 & \mbox{    with probability        }\left( 1+exp(+\frac{h_i(t)}{T})\right)
^{-1}
\end{array}
\right.
\label{probability}
\end{equation}
where all the neurons are updated simultaneously (like in the Little model).
The parameter $T$ measures the noise level of the net and in the noiseless
limit ($T=0$) we recover the deterministic dynamics:
\begin{equation}
S_i(t+1)=Sign \left( h_i^H(t)-\frac \Delta 2(1+S_i(t) \right)
\end{equation}
{}From this expression we can easily understand the effect of this extra field:
if the neuron $i$ fires a spike at time $t$ ($S_{i}(t)=+1$), it will
requires an extra contribution $\Delta$ to the PSP in order to fire again.
On the other hand, if this neuron was at rest at time $t$ ($S_{i}(t)=-1$),
then it will work like an usual Hopfield neuron. Observe that this model
does not distinguish between absolute and relative periods neither includes
any fatigue like effect (long time history). As usual, we will
characterize the recognition ability by calculating the long time behavior
of the overlap $m_{\mu}$ between the state of the system $\{S_i(t)\}$ and
the stored patterns, defined as:
\begin{equation}
m_{\mu}(t) = \frac{1}{N} \sum_{i=1}^{N} \; \xi_{i}^{\mu}\; \langle S_{i}(t)
\rangle_{T}
\end{equation}
where $\langle \ldots \rangle$ means a thermal average at
temperature $T$. We say that the system recognizes a pattern every
time it evolves to an attractor for which only one overlap is
non-zero and all the others vanish as (${\cal O}(1/\sqrt(N))$).
The two relevant parameters in our model are then $\Delta $ and $%
\alpha $ (the ratio between the number of stored patterns ($p$)
and the total number of neurons of the network ($N$)). In the
following sections we analyze the behavior of the model on the
($\Delta $,$\alpha $) plane.

\section{Fixed Point Equation}

\label{Fixed} 

Following the statistical method developed by Geszti
\cite{Geszti} (see also \cite{Peretto,Hertz,Zertuche,Marcus}),
we give in this section  a heuristic derivation of the critical
capacity $\alpha_{c}$ as a function of the parameter $\Delta$
for the stochastic version of the model.
By taking the limit $T \rightarrow 0 $ we obtain a noiseless phase diagram
in the ($\Delta $,$\alpha $) plane which will be compared with numerical
simulations in the next section.

Let us suppose that the initial state of the system is such that $m_1=m$ is
the only macroscopically non-zero overlap and so $m_\mu={\cal O}(1/\sqrt(N))$
for any $\mu \neq 1$. Furthermore, we will assume that although the threshold
tends to destabilize the fixed point attractors, its effect is not strong
enough to anable the system to visit different basins of
attractors. So, since initially only the first overlap was non
zero, let us suppose that this will be valid for any time $t$.
This a priori assumption will be justified in the next section
by the numerical simulation, where we will find that in the
region where the system recognizes (that is, where $m=m_1
(t\rightarrow \infty )\neq 0$) the dynamics of the model is
dominated by fixed point attractors.

We then start considering the overlap between the state of the system and
the first pattern, that can be rewritten as:
\begin{equation}
m=\frac 1N\sum_{i=1}^Ntanh\left( \beta \left( m+\sum_{\mu =2}^p\xi
_i^\mu \xi _i^1m_\mu-\xi _i^1\frac \Delta 2(1+S_i)\right) \right)
\;\;.
\end{equation}
Since we are storing an extensive number of pattern, we cannot neglect
any more the effect of the others $(p-1)$ overlaps:
\begin{equation}
m_\nu =\frac 1N\sum_{i=1}^N\xi _i^\nu \xi _i^1tanh\left( \beta \left(
m+\xi _i^\nu \xi _i^1m_\nu +\sum_{\mu \neq 1,\nu }^p\xi _i^\mu \xi
_i^1m_\mu -\xi _i^1\frac \Delta 2(1+S_i)\right) \right) \;\;.
\end{equation}

In order to make an self-consistent treatment for the overlap $m$ we need to
introduce two other parameters, namely:
\begin{eqnarray}
q  & = & \frac{1}{N} \sum_{i=1}^{N} \langle S_{i} \rangle^{2} \\
 & & \nonumber\\
r  & = & \frac{1}{\alpha} \sum_{nu \neq 1}^{p} m_{\nu}^{2}
\end{eqnarray}
where $q$ is the Edwards-Anderson order parameter and $r$ is
indentified as the mean square overlap of the system
configuration with the nonretrieved patterns \cite{Hertz}.

After some standard calculations we get the following set of equation for
the values of $m$, $q$ and $r$ {\em in the attractor}:
\begin{eqnarray}
m & = & \frac{1}{2} \int Dz \left( tanh\left(\beta L_{z}^{+}\right) +
        tanh\left(\beta L_{z}^{-}\right) \right) \\
 & & \nonumber\\
q & = & \frac{1}{2} \int Dz \left( tanh^{2}\left(\beta L_{z}^{+}\right) +
        tanh^{2}\left(\beta L_{z}^{-}\right) \right) \\
 & & \nonumber\\
r & = &  \frac{q}{\left(1-\beta (1-q) \right)^{2}} \;\; ,
\end{eqnarray}
where $L_z^{\pm }=(1-\frac \Delta 2)m\pm \frac \Delta 2+\sqrt{\alpha r}z$
and
$$
Dz=\frac{dz\exp {-z^2/2}}{\sqrt{2\pi }}\;\;.
$$

Notice that for the particular case $\Delta =0$ we recover the equations
obtained for the Hopfield model \cite{Hertz} which also agree with those
obtained by Amit et al \cite{Amit3} through a thermodynamical mean-field
study (which unlike this method requires the use of the replica trick).

We start analyzing the noiseless case $(T=0)$ for which we have performed
numerical simulations. In this limit our equations take the following form:
\begin{eqnarray}
m & = &  \frac{1}{2} erf\left( \frac{ (1-\frac{\Delta}{2}) m +
         \frac{\Delta}{2} }{ \sqrt{2 \alpha r} } \right) +
         \frac{1}{2} erf\left( \frac{ (1-\frac{\Delta}{2}) m -
         \frac{\Delta}{2} }{ \sqrt{2 \alpha r} } \right)  \\
& & \nonumber\\
C & = &  \frac{1}{\sqrt{2 \pi \alpha r}} \left(
         exp \left( - \frac{ ( (1-\frac{\Delta}{2}) m +
         \frac{\Delta}{2} )^{2} }{ 2 \alpha r } \right) +
         exp \left( - \frac{ ( (1-\frac{\Delta}{2}) m -
         \frac{\Delta}{2} )^{2} }{ 2 \alpha r } \right)
	 \right) \\
 & & \nonumber\\
r & =  & \frac{1}{ (1-C)^{2} } \;\;.
\end{eqnarray}
In Fig. 1 we display $m$ as function of $\alpha $ for several values of $%
\Delta $. For any value of $\Delta <\Delta _c=1$ there always exists a
critical value $\alpha _c$ below which the system recovers the stored
patterns with a non-zero fraction of errors $\epsilon$. At $\alpha _c(\Delta
)$ the systems undergoes a discontinuos transition from the retrieval phase
(in which the dynamics is governed by the fixed point attractors) to a
non-retrieval phase where our analytical approach is no longer valid,
since the self-consistent equation does not predict a fixed point
attractor (which was our original assumption). Observe that $\alpha _c$
decreases as $\Delta $ increases. As $\alpha \rightarrow 0$ the fraction of
errors at the transition $\epsilon_{c}=\frac 12(1-m)$ goes to $0$
accordingly to the following expression:
\begin{equation}
\epsilon=\frac 12\sqrt{\frac \alpha {2\pi }}\left( exp(-\frac 1{2\alpha })+
\frac{exp(-\frac{(1-\Delta )^2}{2\alpha })}{1-\Delta }\right)
\end{equation}
We have also analyzed the fixed point equations in the presence of noise. In
Fig. 2 we present the $(T,\alpha)$ phase diagram for different values
of $\Delta$. For $\Delta =0$ we recover the phase diagram
obtained in \cite{Amit3}. Along the lines $T_{c}(\alpha )$ the
system undergoes a discontinuos transition from the retrieval
phase (below) to the non-retrieval phase (above). Notice that the
recognition phase decreases as $ \Delta $ increases; i.e., the
main effect of introducing this refractory periods seems to be a
degradation of the retrieval properties of the model.
In Fig. 3 we present the critical line $T$ versus $\Delta$ for
$\alpha =0$.  For $\Delta < 0.611$ the system undergoes a second
order transition while for $\Delta > 0.611$ the transition is
discontinuos (the point $(T,\Delta)\simeq (0.46,0.611)$ separates
both lines). In the inset we show the behavior the
retrieval overlap around the critical point as function of $T$.

%

\section{Numerical Simulation}

\label{Numerical} 

In this section we present a numerical study of both recognition ability and
dynamical properties of the model at $T=0$ and compare it with the
analitical results obtained in the previous section. The simulations were
performed on systems of $N=800$, $1600$ and $3200$ neurons and the network
was updated synchronously. Setting the initial configuration as the first
stored pattern, we let the system evolve until it reaches the attractor.

In order to characterize the dynamical behavior we first determined whether
the system was in a periodic orbit or not, by waiting until it returned to a
given configuration that was stored after a transient. Depending on the
value of the parameters and on the size of the system it could also happen
that the system did not return to the initial configuration after a given
period of time (typically 100 Monte Carlo Steps). In such cases, we said that
the system follows a chaotic orbit, although we have not
performed a through analysis in order to determine whether these
were really chaotic orbits or orbits with large periods.

To analyze the recognition ability we calculated for each sample a temporal
average between the stored patterns and the state of the system in the
attractor. If the system reached a cyclic orbit of period $t_c$, we measured
(in the attractor) the following quantity:
\begin{equation}
m_\mu =\frac 1{t_c}\sum_{t=t_0}^{t_0+t_c}m_\mu (t)
\end{equation}
Since the initial state was chosen to be always the first memory, we say
that the network recognizes when
\begin{equation}
m=m_1\sim {1}\hspace{2cm}m_\mu \sim {\cal
O}(1/\sqrt(N)),\hspace{0.2cm}for \hspace{0.1cm}\mu >1
\end{equation}

In order to make a configurational average of $m$, for any value of the
parameters we repeated this procedure over $100$ different samples using
different memories, initial configurations and random number sequences. To
characterize the dynamical behavior we present the frequency with which
each kind of attractor appears and also the mean activity,
defined as the average number of active neurons, in the attractor.
\begin{equation}
a =\frac 1{2N}\sum_{i=1}^{N}(1+S_{i})
\end{equation}

In Fig. 4 we display the phase diagram $\Delta$ vs. $\alpha$
for $N=3200$. For $\Delta = 0$ the system presents only fixed
points (FP). For fixed $\alpha$, as $\Delta$ increases we found that:

\begin{enumerate}
\item  for low values of $\Delta $ the dynamics is governed only by fixed
points attractors. The full circle indicates where this kind of behavior
disappears;

\item  the region between the two full triangles indicate the region where
cycles of order two (C2) appear;

\item  the hollow circle indicates the value of $\Delta $ above which
chaotic orbit (Ch) emerges.
\end{enumerate}

Observe that there are many region of coexistence of attractors. In fact,
between the C2 and the Ch we have also found cyclic orbits (OC)
of order greater than two, but they are not indicated in the diagram.

Independently of the dynamical behavior, we have also studied the critical
recognition capacity. The dashed line separates the recognition phase
(below) from the non-retrieval phase (above) obtained numerically and the
full line corresponds to the analytical results obtained in the previous
section. The simulation curve fits very well the analytical result only for
small values of $\alpha $.

In order to understand why the analytical and the numerical curves do not
agree, we have carefully analyzed the behavior of the system along two cuts
with fixed $\alpha $, namely $0.01$ and $0.04$. In Fig. 5 we plot both $m$
(top) and the frequency with which each kind of orbits appears (bottom) as a
function of $\Delta $. The first thing we note is that the FP
region coincides with the retrieval phase, and that the C2 region
corresponds to the non-retrieval phase. In such cases, where the
systems only recognizes through FP, the analytical curve
predicts very well the transition. On the other hand, in Fig. 6 we present
the same curves for $\alpha =0.04$. Notice now that the recognition phase
presents two different dynamical behaviors: for small values of $\Delta $ the
system evolves to FP while for intermediate values it goes to
C2. Unlike the $\alpha =0.01$ case, now the theoretical curve
does not predict correctly the retrieval non-retrieval phase transition, but
the FP to C2 transition. In both cases we have studied the
finite-size effects by working with three different sizes, namely, $N=800$, $%
1600$ and $3200$. In Figs. 5 and 6 we present the overlaps as function of $%
\Delta $ for all these system sizes. Note that as $N$ increases the
numerical simulation tends to display a more abrupt decay of $m$ at the
transition, resembling the first order transition found in the analytical
calculation.

Finally, in Fig. 7 we show the mean activity as function of
$\Delta$ for $\alpha = 0.01$ and $0.04$. We can notice that where
there are fixed points and periodic orbits with recognition
within, the mean activity remains around the value $\sim 0.5$ (random
variables) and it only decreases in the transition to
non-retrieval phase. This shows that the parameter $\Delta$ not only
damages the recognition ability but also destabilizes the
tendency of the system to evolve to fixed point attractors,
allowing the appearence of more complicated retrieval attractors.


\section{CONCLUSIONS}

\label{conclution} 

In this work we study analytically and through of numerical simulations a
model for associative memory where we have incorporated in the dynamics of the
network a new kind of threshold that simulate the effect of the
refractory period.
The main result is that the parameter $\Delta $ that activates
this threshold yields to the appearing of Chaotic and periodic
attractors. Nevertheless, the system seems to recognizes only through
fixed point and cycles of order two. Only in a small region the
system recognizes with higher order cycles and with chaotic
trajectories, but this behavior appears just in the boundary
between the retrieval and the non-retrieval phases. It would be
interesting to make a more detailed study to elucidate whether this kind
of trajectories are due to finite size effects or not. As much
as we could see, as $N$ increases they do not seem to
dissapear, so we suspect that they will exist also in the
thermodynamical limit.

In the recognition phase (small values of $\Delta$), the PSP is strong
enough to drive the system to stable attractors, FP and
periodic orbits, where the average overlap in each regime is of
the order $1$. For large values of $\Delta $ the performance is
drastically damaged, and in these regions  the dynamics is
dominated by very large cycles or chaotic trajectories.

The numerical simulation fits very well the analytical results
only for small values of $\alpha$, where the transition occur
from fixed point FP to cycle order two C2. Actually, the
analytical curve seems to fit only the line where the fixed
point behavior disappears. We also observe that in the
transition the mean activity decreases with the increase of
$\Delta$.

\vskip 3\baselineskip
{\bf Acknowledgements:} We acknowledge to D. A. Stariolo and
F. S. de Menezes for fruitful
discussion. We thank the Supercomputing Center of the Universidade
Federal do Rio Grande do Sul (CESUP-UFRGS) for use of the Cray
YMP-2E. This work was supported by Brazilian agencies CNPq and
FINEP.
\newpage
%
%
%
\noindent {\bf CAPTIONS FOR FIGURES}

\vspace{1cm}
{\bf Figure 1.} Plot of $m$ versus $\alpha$ at $T=0$ for different
values of $\Delta$. At $\alpha_{c}(\Delta)$ the system undergoes
a discontinuous transition from the recognition phase to
non-retrieval phase.\\

{\bf Figure 2.} Phase diagram $T$ versus $\alpha$ for $\Delta =
0,0.2,0.4$ and $0.6$. Below of the critical lines the system
recognizes with fixed point and the transition to non-retrieval
phase is discontinuos.\\

{\bf Figure 3.} The critical line $T=f(\Delta)$ for $\alpha=0$.
For $\Delta < 0.611$ the transition is of second order (full
line) while for $\Delta > 0.611$ the transition is discontinuos
(dashed line). $(T,\Delta)\simeq (0.46,0.611)$ is a critical point.\\

{\bf Figure 4.} The numerical phase diagram $\Delta$ versus
$\alpha$ at $T=0$ and $N=3200$, showing the regions FP (below full
circles), periodic (between the two full triangles) and Ch (above hollow
circles). The simulation (dashed line) and analytical (full
line) curves separetes the recognition phase (below) from
non-retrieval phase (above).\\

{\bf Figure 5.} Plot of $m$ (top) and of the frequency (bottom) in
the which each kind of orbits appears as a function of $\Delta$
for $\alpha=0.01$. The full line corresponds to the analytical curve.\\

{\bf Figure 6.} Plot of $m$ (top) and of the frequency (bottom)
in the which each kind of orbits appears as a function of
$\Delta$ for $\alpha=0.04$. The full line corresponds to the analytical
curve.\\

{\bf Figure 7.} The mean activity vs. $\Delta$ for $\alpha=0.01$
and $\alpha=0.04$.

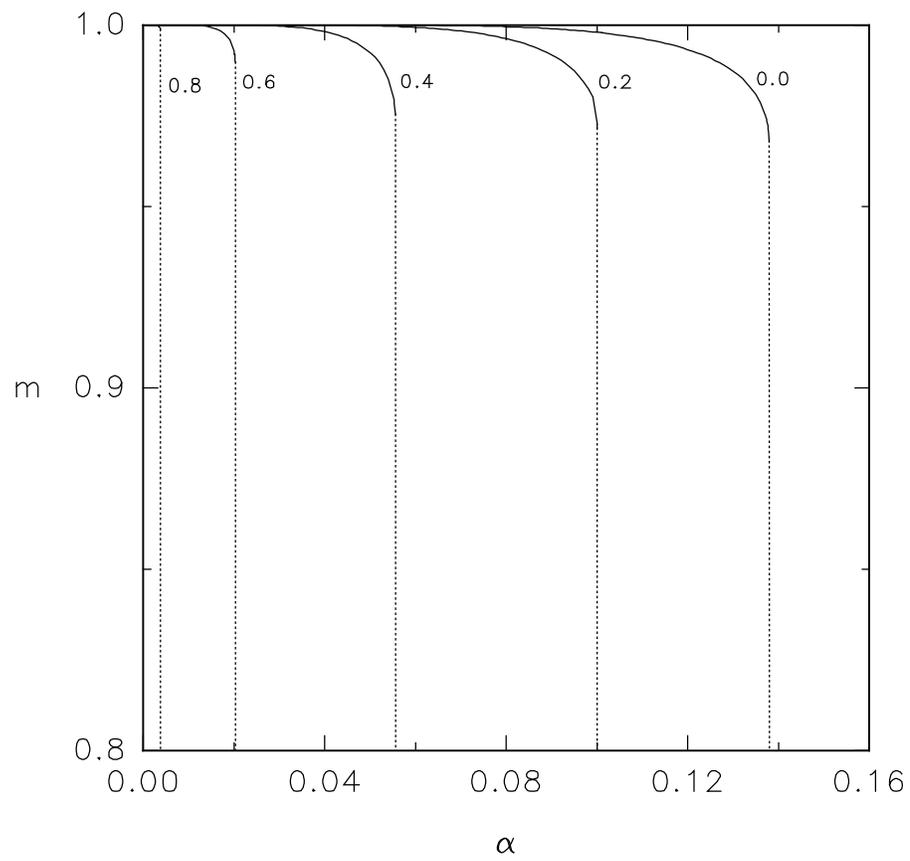
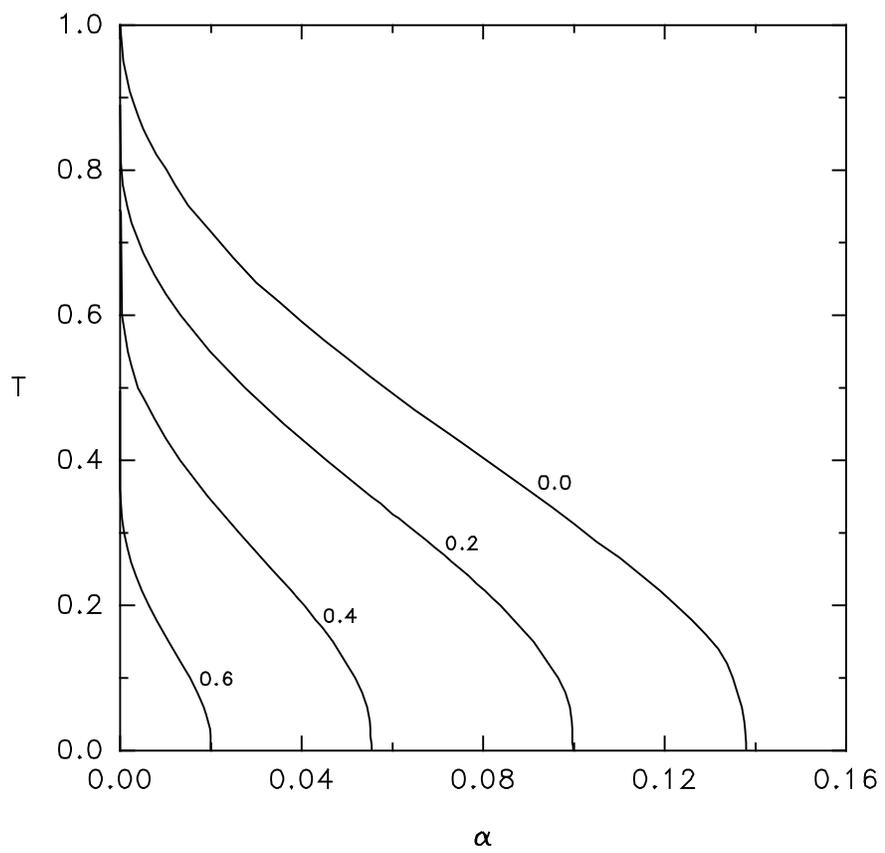
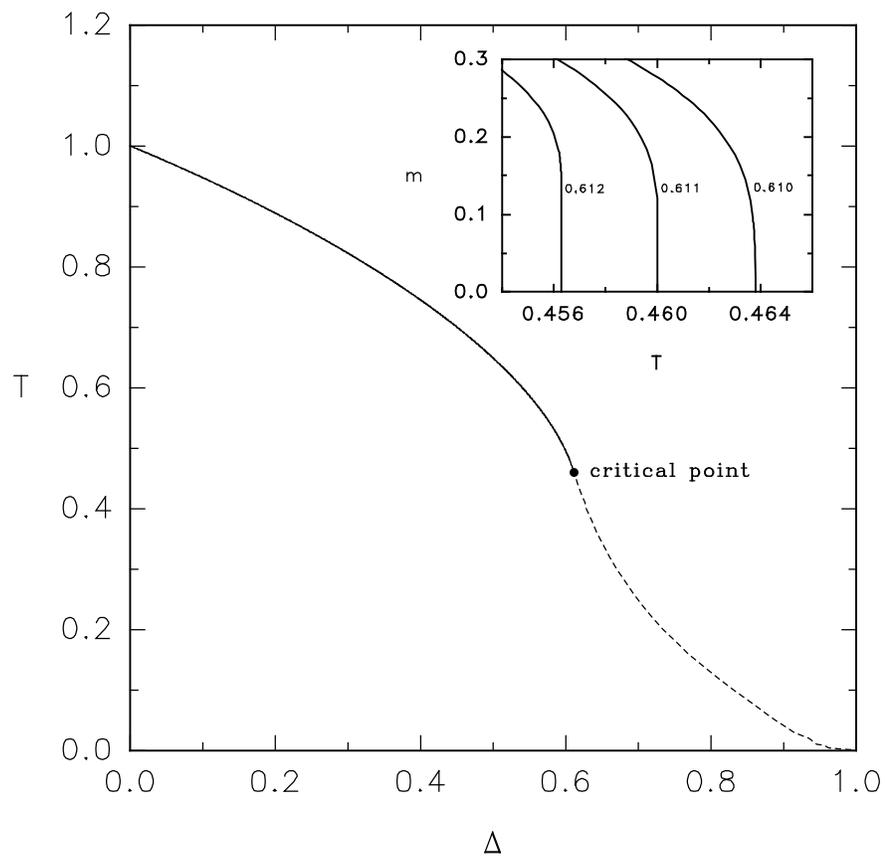
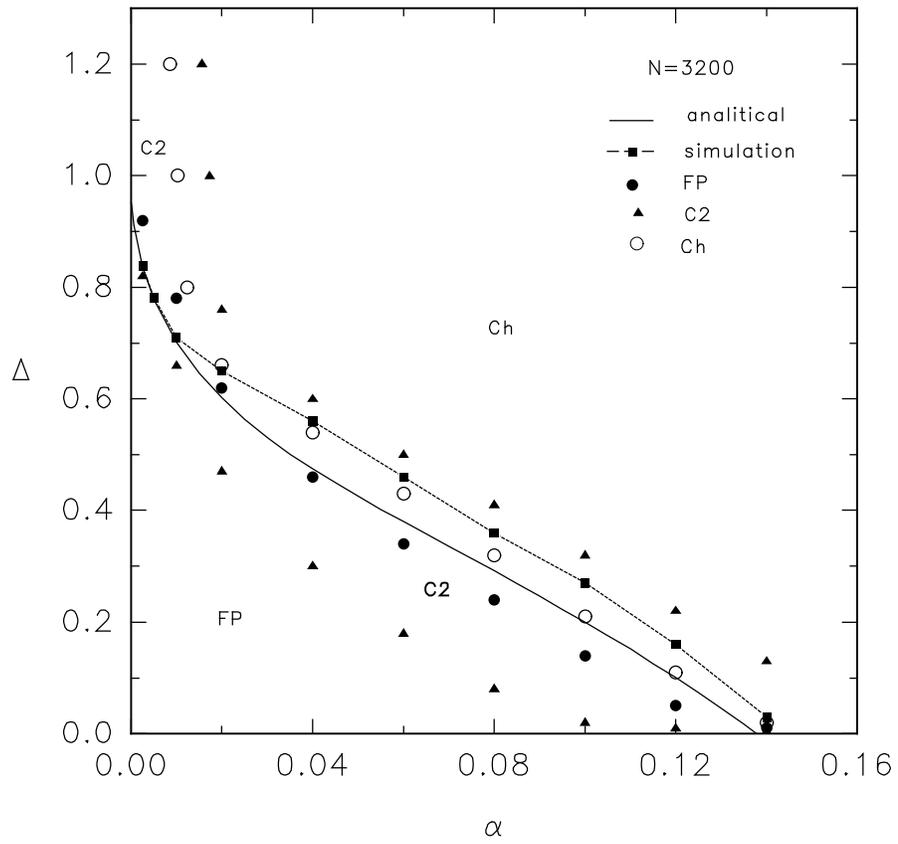
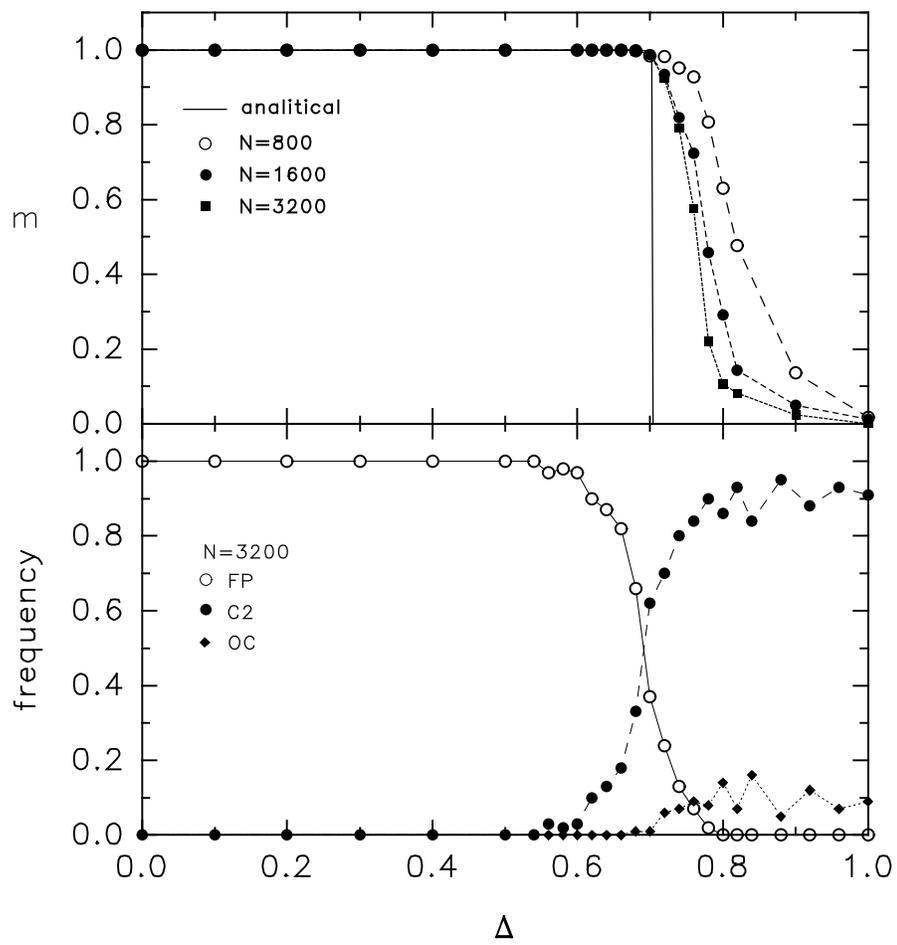
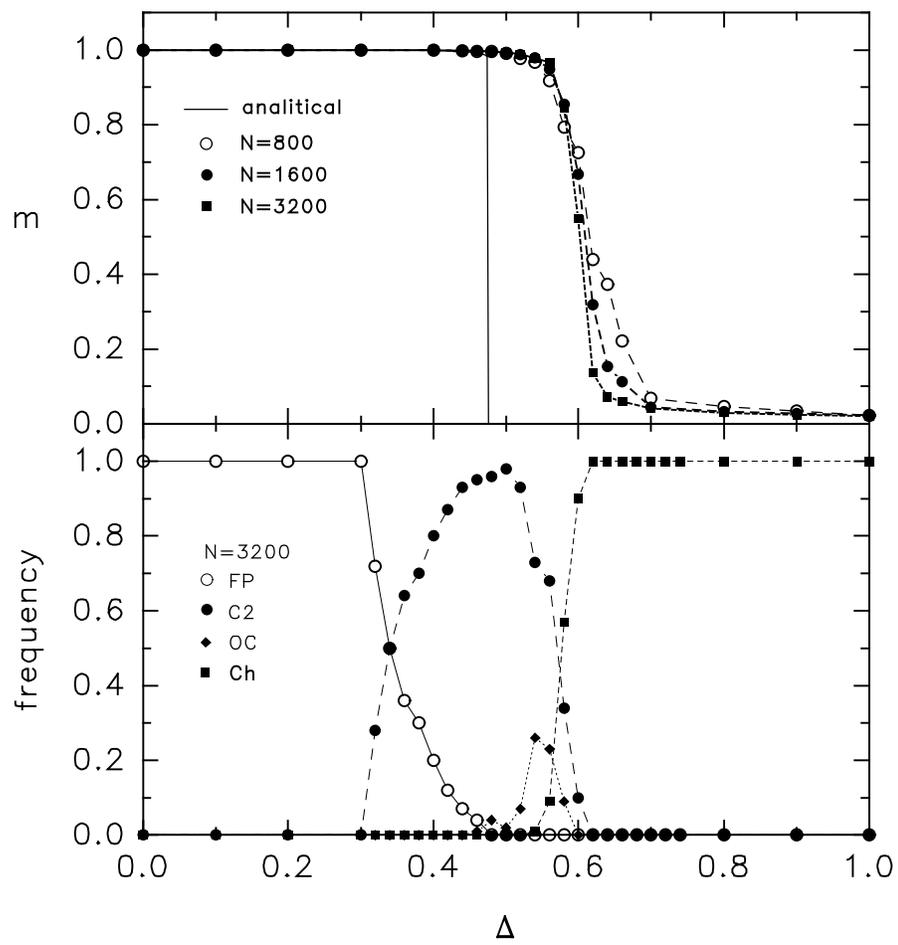
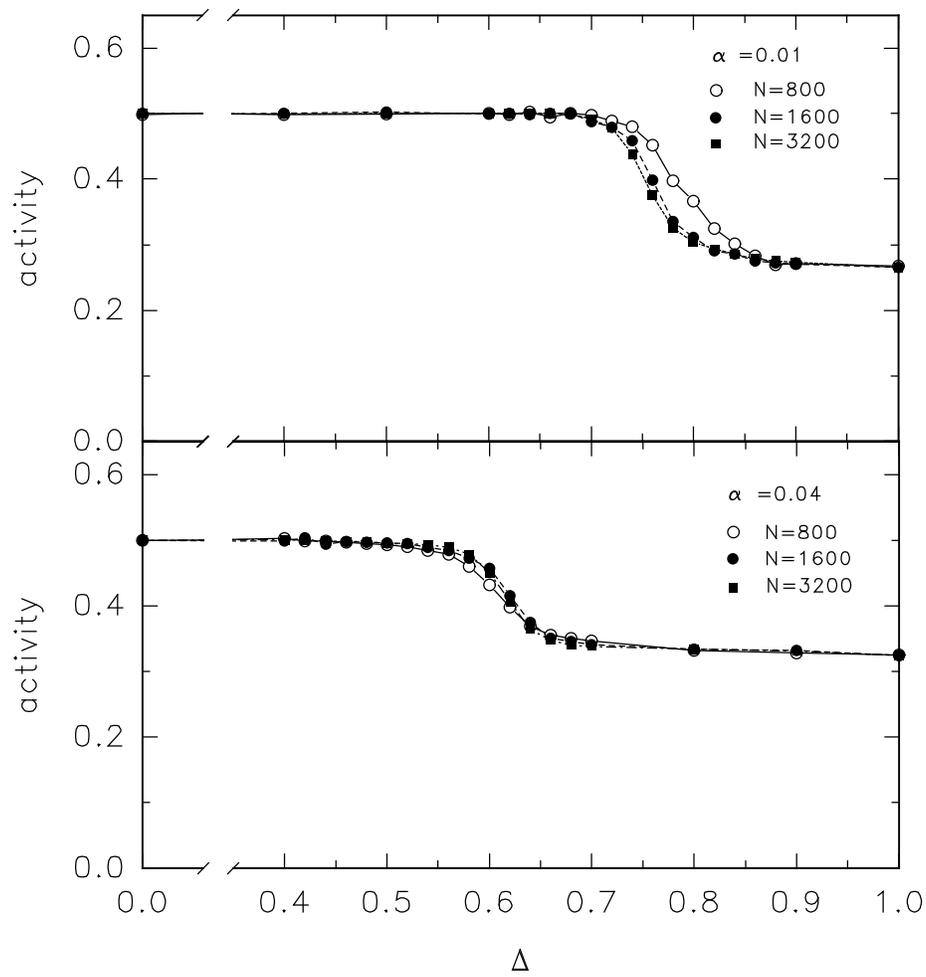
\end{document}